# Ambient condition superconductivity via engineered polaronic environment

Krzysztof Kempa and Michael J. Naughton
Department of Physics, Boston College, Chestnut Hill, Massachusetts 02467, USA

**ABSTRACT**

A vanishing dielectric function is required for longitudinal plasmonic or polaronic modes in a polarizable uniform medium, and heralds the presence of singular charge fluctuations. It also strengthens Cooper pairing in a superconductor via resonant anti-shielding (RAS). We combine these notions in a strategy to strongly enhance superconductivity. Specifically, we propose a superlattice of an ultrathin superconductor and a metal-organic framework material. This structure possesses a momentum-independent and resonant effective dielectric function, a key feature for RAS-enhanced superconductivity. We show that the superlattice facilitates near perfect volumetric intermixing between the superconductor and the engineered dielectric environment. To estimate the $T_c$ enhancement, we use the Leavens scaling method, which relates the spectral integral of the RAS-renormalized Eliashberg function $\alpha^2F$ to $T_c$, which we showed previously to be in excellent agreement with *ab initio* simulations, and which we posit may underlie recent observations of enhanced $T_c$ in FeSe on $SrTiO_3$. After renormalizing the bare $\alpha^2F$, dividing it directly by $|\varepsilon_{DE}(\omega)|^2$, where $\varepsilon_{DE}(\omega)$ is the $q$-independent dielectric function of the dielectric environment, our estimates show that ambient temperature and pressure operation could be achievable in the RAS scheme. To avoid lattice reconstruction or heavy bipolarons, both deleterious to superconductivity, we propose a system that assures that the corresponding electron-boson coupling remains moderate ($\lambda$ of order 1). We also calculate the quantum Fisher information from the dynamic charge susceptibility in the normal state (for vanishing losses) in our proposed superlattice, and our estimate suggests the presence of superconductivity signatures well above $T_c$ due to quantum charge entanglement buildup.

kempa@bc.edu https://orcid.org/0000-0003-3953-1485
naughton@bc.edu https://orcid.org/0000-0002-6733-2398

A superconductor that operates at ambient conditions temperature and pressure has eluded researchers so far. In the 1970s, Ginzburg and Kirzhnits [1] argued that there are no physical limits to ambient operation of BCS superconductors. Some fifty years later, however, the superconductivity record achieved in the cuprates in 1993 still holds, with $T_c \sim 133$ K [2]. Equally disappointing has been theoretical progress, with the origin of superconductivity in non-BCS superconductors (including cuprate and possibly other families) remaining unknown or, at best, unclear. Consequently, calculations of $T_c$ for a given structure are also unreliable, which in turn makes any attempt to reverse-engineer the problem impossible.

A potential alternative to discovering a new superconductor is to engineer an existing one's dielectric environment (DE). In fact, the possibility of a singular enhancement of $T_c$ via DE engineering was evident in early (1950-60) work by Migdal [3] and Eliashberg [4]. Later, Kirzhnits, *et al.* [5] reformulated BCS in the language of a dielectric function. In the diagrammatic representation, the ω- and $q$-dependent dielectric function $\varepsilon(\omega, q)$ enters the electron-phonon interaction diagram (*i.e.* the Frohlich term) in the form $|\varepsilon(\omega, q)|^{-2}$, making it (and $T_c$) singular for $\varepsilon(\omega, q) \to 0$ [6]. This is also the well-known condition for the existence of longitudinal plasmonic modes in a uniform medium, and in general, it indicates the presence of singular charge fluctuations in a dielectric polarizable system. We call this here a resonant anti-shielding (RAS) effect, since it occurs for frequencies above the resonance, and thus involves a phase change of the charge fluctuations reversing the usual shielding effect of a DE. Note that this $T_c$ enhancement strategy via RAS is universal, since it relies on the fact that Cooper pairs, regardless of their pairing mechanism, are simply charges that can be affected by external *E*-fields induced in an engineered dynamic DE.

Achieving RAS in practice is quite challenging, however, requiring the following stringent conditions to be simultaneously satisfied: (1) RAS must occur for frequencies in the range of the pairing excitations; (2) it must occur in the entire range of momentum transfer wave vectors $q$, specific for a given superconductor; (3) $\varepsilon(\omega, q)$ must be well-defined, amenable to a mean field description of the DE; and finally (4), RAS must not lead to excessive enhancement of the electron-phonon coupling, *i.e.* λ must remain of order 1. The 2nd condition can be achieved only when $\frac{d\varepsilon(\omega,q)}{dq} \to 0$, *i.e.* when $\varepsilon(\omega, q)$ is $q$-independent, which dramatically enhances the density of states (DOS) of the RAS-capable states. To prove that, note that RAS occurs for a well-defined $\omega = f(q)$, a dispersion relation of the corresponding polarization mode, and so (in 3D) DOS $= \frac{dN}{d\omega} \sim \frac{dN}{dq}\frac{dq}{d\omega} \sim q^2 \frac{dq}{d\omega} \sim \left(\frac{d\omega}{dq}\right)^{-1}$, which diverges when $\frac{d\omega}{dq} \to 0$ , *i.e.* in the absence of dispersion for a mode corresponding to RAS. Finally, this also implies that $\frac{d\varepsilon(\omega,q)}{dq} = \frac{d\varepsilon(\omega,q)}{d\omega}\frac{d\omega}{dq} \to 0$, which completes the proof. This could be understood also as a result of the *N*-fold degeneracy of $\varepsilon(\omega, q)$ with respect to the variable $q$; there are $N \to \infty$ RAS conditions for a single $\omega$, as opposed to a single RAS condition for a single $\omega$ if $\varepsilon(\omega, q)$ is $q$-dependent, and thus a much more robust effect. The 4th condition results from the fact that at such an excessive enhancement of the electron-phonon coupling, there is an increasing chance of lattice reconstruction or formation of heavy bipolarons, deleterious to superconductivity [7]. On the other hand, a superfluid condensate of light-weight bipolarons, at moderate electron-phonon coupling, can strongly enhance the superconductivity [8].

Attempts to obtain RAS by employing metamaterials has led to only small $T_c$ enhancements [9,10]. Such materials are simply unable to satisfy the above conditions. However, recent advances in 2D superconductors, in particular the 10-fold $T_c$ enhancement in monolayer FeSe on strontium titanate (STO) [11], provide direct evidence for a strong DE enhancement consistent with our RAS theory. Specifically, it has been demonstrated that collective plasmon-polaron modes of STO, which of course herald the presence of RAS, are weekly dispersive [12], and strongly interact with Cooper pairs in the FeSe monolayer [13]. We have recently proposed [14] an explanation for the 10-fold $T_c$ enhancement of $T_c$ in the FeSe/STO system by demonstrating that the RAS effect triggers a dramatic transformation of the broadband Eliashberg function in the absence of RAS $(\alpha^2 F)$ into a single peak. This is consistent with the bipolaron picture of Bose-Einstein-like condensation. We demonstrate this effect here as well.

In recent work [15], we investigated also the topological crystal $Bi_2Se_3$, which supports an exotic plasmonic mode, a hybrid of a topological plasmon and a phonon [16]. We demonstrated that this mode leads to a dielectric function form that satisfies RAS in a narrow frequency-momentum range, and thus could lead to a significant enhancement of superconductivity. Here, we propose a potentially ideal system for inducing RAS: the point-dipole crystal (PDC). To achieve strong $T_c$ enhancement with a PDC, we propose a superlattice arrangement with a unit cell consisting of a superconductor monolayer in direct contact with an ultrathin PDC. We also propose a metal-organic framework (MOF) as an excellent proxy for a PDC, and show that RAS in the analogous MOF-superconductor superlattice can also strongly enhance $T_c$.

The PDC was proposed by Mahan and Obermair [17] as a model for the electromagnetic response of an atomic crystal. It consisted of a cubic array of point-dipoles, and its response was obtained using a fully retarded treatment in the Coulomb gauge. The full dynamic electromagnetic response of this PDC for polarizable and harmonically resonant point dipoles was carried out by Kempa, Rupin and Pendry [18], who showed that the PDC is versatile, and allows for bulk and surface plasmon wave propagation, as well as negative refraction in a polaritonic band. An extension of this work [19] showed that a PDC has its dielectric function *independent of* $q$, essentially in the entire Brillouin zone, which has a simple Lyddane-Sachs-Teller (LST) form [20]

$$\varepsilon(\omega, q) = \varepsilon(\omega) = \frac{\omega_{LO}^2 - \omega(\omega + i\gamma)}{\omega_{TO}^2 - \omega(\omega + i\gamma)}, \tag{1}$$

with $\omega_{TO}$, $\omega_{LO}$ characteristic frequencies and $\gamma$ a loss factor. It is important to note that this $q$-independent form is exact, and not the result of a long wavelength limit. Physically, this is because of the phonon modes of atomic size ($\approx$ 0.3 nm) oscillators are highly localized, which translates into $q$-independent dispersion, at least for $q$ of order $1/a \approx 3\ \text{nm}^{-1} \gg k_F$.

As mentioned, an excellent proxy for PDCs could be obtained with MOF systems. In one example, a 2D MOF was developed with a pair of Mo atoms as the polarizable unit, square-periodically distributed in a planar organic matrix [21]. Raman spectra show a sharp feature near 380 cm$^{-1}$ ($\hbar\omega \approx 47$ meV), which represents highly localized phonon oscillations of the Mo atom pair [21]. In addition, due to atomically-sized oscillating dipoles (the Mo atoms), and since other resonant terms in the response vanish at large required values of $q$, the MOF structure is indeed an excellent proxy for a PDC, with the dielectric function formula accurately described by Eq. (1), with $\omega_{TO} = 47.5$ meV and $\gamma = 1.32$ meV. From the experimental and simulation data on similar MOFs [22], we deduce $\varepsilon_{MOF}(0) \approx 2$ and thus from Eq. (1), we get the plasmon depolarized at $\omega_{LO} \approx 67$ meV, very close to the maximum of the Eliashberg function for the superconductor $MgB_2$.

There are additional resonances near the $\omega_{TO} = 47.5$ meV resonance of interest in the MOF spectrum, which result mostly from the phonon activity in the matrix [21,22]. These can be easily accounted for by using the large $q$ dielectric function extension [23], which leads to a simple hydrodynamic form for the $q$-dependence of the other terms:

$$\pi_{ot}(\omega, q) = \sum_{n=1}^{M} \frac{\omega_p^2}{\omega_n^2 + \beta q^2 - \omega(\omega + i\gamma_n)} \tag{2}$$

where $\omega_{pn}^2$ is the oscillator strength of the $n^{\text{th}}$ term, $\omega_n$ is the frequency of the other resonances, and $\gamma_n$ is the corresponding loss factor. For a MOF at the required very large $q$, $\pi_{ot}(\omega_{TO}, q) \ll 1$, as long as these terms are due to the usual dispersive phonon modes. Thus, these contributions to the large $q$ response are vanishing, and so $\varepsilon_{MOF}(\omega, q)$ is indeed dominated by the resonance at $\omega_{TO} = 47.5$ meV. This is an unexpected benefit of the required large $q$'s.

In order to proceed with the dielectric function formalism, a composite material must first have a well-defined effective $\varepsilon(\omega, q)$. Consider a system of small, dynamically-active, polarizable elements of size $a$, immersed in a passive matrix, spaced a distance $d \approx a$ apart. The most important dimensional scale here is the superconducting coherence length $\xi$; therefore, the condition for well-defined effective $\varepsilon(\omega, q)$ is simply $d \ll \xi$. In practice, this condition is challenging to satisfy with good superconductors, where $\xi$ is very small ($\xi < 10$ nm). This requires $d$ to approach atomic scale, effectively eliminating conventional metamaterial strategies. This, however can readily be achieved using MOF technology, and the superlattice structure. Such a structure, with the MOF of Ref. [21] sandwiching unit cells of $MgB_2$, is shown schematically in Figure 1(a). Since $d \approx 0.35$ nm (greater than any other elemental polarizable entities/atoms in this configuration, and equal to the layer thickness for both $MgB_2$ and MOF), the inequality above is well satisfied, not only for $MgB_2$ at $T_c$ = 39 K ( $\xi \approx 7$ nm [24]), but also should be approximately satisfied even when the enhanced $T_c$ could approach room temperature, where $\xi$ is expected to decrease to 1.5 nm [25]. As shown above, the frequency part of the RAS condition is also satisfied for this superconductor, and so is the large $q$-requirement, since the corresponding dielectric function is $q$-independent almost in the entire Brillouin zone, *i.e.* $\varepsilon_{MOF}(\omega, q) \approx \varepsilon_{MOF}(\omega)$.

A schematic of the proposed structure is shown in Figure 1, where the superconductor is effectively immersed in the DE of the MOF, uniformly distributed inside the superconductor, as a superlattice of unit cell-thick layers of MOF ($d \approx 0.35$ nm) and $MgB_2$ ($D \approx 0.35$ nm). As such, a simple Maxwell-Garnet effective medium averaging [26] can be used to provide a rough estimate of the effective $\varepsilon$ of the DE, given by

$$\varepsilon_{DE}(\omega) = (1 - f) + f\varepsilon_{MOF}(\omega) \quad (3)$$

where we assume that $\varepsilon$ of the superconductor itself is 1 (as expected for very large $q$), and the MOF filling parameter is $f = \bar{d}/(D + \bar{d})$, where $\bar{d} \approx de^{-qd}$. Since $D = d$, and assuming that maximum $q \approx 1/d$, we expect $f \leq 0.3$. We emphasize that this superlattice, combined with the proximity effect, mimics the bulk effect by perfectly mixing the superconductor with the DE. This is the most important advantage of the superlattice geometry.

A reliable estimate of $T_c$ for superconductors can be obtained by employing the scaling method of Leavens [27]. As was demonstrated in our earlier paper [15], this method, in spite of its simplicity produces good agreement with an *ab initio* simulation employing numerical solutions to coupled Eliashberg equations. We employ this method in our exploratory work here as well. First, one must know the Eliashberg function $\alpha^2F(\omega)$ for a given superconductor. Figure 1(b) shows $\alpha^2F(\omega)$ *ab initio* simulated for $MgB_2$ [15, 28]. Leavens' method estimates not $T_c$ but its least upper bound $T_c^{max}$, which is a simple frequency functional [27]

$$T_c^{max} = c\int_0^\infty \alpha^2F(\omega)d\omega \quad (4)$$

where $c$ is a slowly varying function of the pseudopotential, and ranges from 0.15 – 0.23 [26]. Using the $\alpha^2F$ for $MgB_2$ we get $40\ K < T_c^{max} < 60K$, in the range of experimental values for this material. Next, we calculate the RAS-renormalized Eliashberg function [6,28], which is given by

$$\overline{\alpha^2F(\omega)} = \frac{\alpha^2F(\omega)}{|\varepsilon_{DE}(\omega)|^2} \approx \frac{\alpha^2F(\omega)}{|(1-f)+f\varepsilon_{MOF}(\omega)|^2}\,, \quad (5)$$

and is shown in Fig. 1(c). Clearly, there is a dramatic transformation of the Eliashberg function, from the broadband spectrum shown in Fig. 1(b) to an essentially single peak shown in Fig. 1(c), in agreement with the bipolaron condensation/superfluidity concept. Finally, the corresponding $T_c$, renormalized by RAS, is given by

$$\overline{T_c^{max}} = c\int_0^\infty \overline{\alpha^2 F(\omega)} d\omega = c\int_0^\infty \frac{\alpha^2 F(\omega)}{|(1-f)+ f\varepsilon_{MOF}(\omega)|^2} d\omega. \tag{6}$$

Eq. (6) applied to this transformed function at $f = 0.2$ (safely below the critical value 0.3 given above) gives. 168 K < $T_c^{max}$ < 258 K. By choosing the critical value $f = 0.3$, on the other hand, Eq. (6) yields 320 K < $T_c^{max}$ < 490 K (room temperature operation). Clearly, increasing the MOF filling parameter $f$ enhances $T_c^{max}$. However, this enhances $\lambda$ as well, since

$$\lambda = 2\int_0^\infty \frac{\alpha^2 \mathrm{F}(\omega)}{\omega} \mathrm{d}\omega \tag{7}$$

In fact, with the delta-function form of $\overline{\alpha^2 F(\omega)}$ (like that shown in Fig. 1(c)), $T_c$ is proportional to $\lambda$ [14]. For $\lambda > \lambda_{max}$ lattice reconstruction and/or formation of heavy bipolarons can occur [7]. However, as discussed in [14], the quantity $\lambda_{max}$ is not well-defined. For example, in YBCO, with $T_c \approx 90$ K (without RAS), the corresponding value is $\lambda = 2.15$, and in the record high $T_c$ cuprates ($T_c \approx 150$ K), $\lambda \approx 3.6$ is expected, and thus $\lambda_{max} > 3.6$. Therefore, the case of the MOF-$MgB_2$ system studied here for $f = 0.2$ (see Fig. 1), for which $\lambda = 3.52$, should be expected to avoid the lattice reconstruction and heavy bipolarons. It is unclear if the room temperature operation above, which occurs in our system for $f = 0.3$, can be achieved, since in this case $\lambda = 6.6$.

The $T_c$ enhancement in our scheme relies on several conditions listed in this work: validity of the effective medium analysis with well defined $\varepsilon(\omega, q) \approx \varepsilon(\omega)$, presence of a robust RAS effect, and extreme proximity between the DE of the MOF and the superconductor. All of these are approximately satisfied in the proposed structure. In particular, proximity is absolutely necessary, and the proposed superlattice arrangement maximizes this effect. It relies also on the superconducting proximity effect, by assuring that the superconducting interlayer distance is much smaller than the coherence length.

Fabrication of such an ultrafine superlattice will require fine control over the chemistry and physics of material-interfacing at the atomic scale, to avoid problems such as lattice mismatch leading to defects or unwanted chemical activity, that in turn might destroy the physics of the ultrathin layers via charge transfer or trapping, material interdiffusion, *etc*. Such efforts are ongoing, and will benefit from guidance by fully *ab initio* simulations, not only of the superconductor, but also of the DE (ground state and time-dependent DFT), to account properly for the structure as well as the response of the DE. We have also explored whether this DE strategy could be applied to the cuprates, and found $T_c$ enhancements similar to that in Figure 1. The present prediction of increasing $T_c$ does not yet consider the anisotropic nature of either the superconducting coherence length or the resonant coupling of Cooper pairs to the MOF-controlled DE. As elaborated, however, we cast the model within the context of these two parameters being sufficiently large in the interlayer distance to facilitate the effect. Thus, while we depict a multilayer structure in the figure, we do not discount the possibility of a measurable effect in a simpler, few-layer structure (such as thin film MOF on thin film superconductor, or *vice versa*), so long as the interlayer coherence length extends across many unit cells. We also point out that a key advantage of using MOFs is the wide tunability of these systems with respect to their unit cell dimensions and choice of metal atoms/clusters that provide the dielectric resonance.

Finally, since RAS enhancement is dominated by spectral characteristics of $\varepsilon_{DE}(\omega)$, or the electronic dynamic susceptibility ($\chi_{DE}(\omega) = \varepsilon_{DE}(\omega) - 1$), we can assess to what degree this is related to charge entanglement in an enhanced superconducting system in the normal state. We can use the relation for quantum Fisher information (QFI) [29] that employs the general Kubo response function

$$F_Q(T) = \frac{4}{\pi}\int_0^\infty d\omega \tanh\left(\frac{\hbar\omega}{2kT}\right) \chi''(\omega, T) \quad (7)$$

where $\chi''(\omega, T)$ is the imaginary part of the dynamic susceptibility in the state $\rho$ with respect to $\hat{\vartheta}$ (the same thermal state and generator for which QFI is evaluated). Assuming that $\chi''(\omega, T) = \mathrm{Im}[\chi_{DE}(\omega)] = \mathrm{Im}[\varepsilon_{DE}(\omega)] = \mathrm{Im}[\varepsilon_{MOF}(\omega)]$, and using Eq. (1), for simplicity with $\gamma \to 0$, we get

$$F_Q(T) = 2f\left(\frac{\omega_{LO}}{\omega_{TO}} - \frac{\omega_{TO}}{\omega_{LO}}\right)\tanh\left(\frac{\hbar\omega_{TO}}{2kT}\right) \quad (8)$$

For a bare superconductor, $f = 0$, which implies $F_Q(T) \approx 0$ in this lossless limit. However, when MOF-induced RAS is present ($f \neq 0$), $\omega_{TO} = 47$ meV, $\omega_{LO} \approx 60$ meV, and $f = 0.5$, and then QFI is nonzero even at the room temperature, $F_Q(300K) \approx 0.5 \tanh\left(\frac{47}{52}\right) = 0.36$, perhaps heralding charge entanglement above $T_c$. This might confirm the possibility of entanglement influencing superconductivity [29]. A universal behavior, characteristic of the quantum critical region, has been observed in a high-$T_c$ superconductor through measurements of optical conductivity, simply related to $\chi''$ [30].

In summary, we proposed a superlattice arrangement of a superconductor in direct contact with a point-dipole crystal, a model system known to possess a momentum-independent effective dielectric function at an adjustable frequency, the key feature for strong plasmonic enhancement of superconductivity. The superlattice arrangement optimizes the desired, near perfect volumetric intermixing between the superconductor and the engineered environment, and a metal-organic framework (MOF) is an excellent proxy for point-dipole crystal. Our estimates of $T_c$ show the potential for ambient temperature and pressure operation in this configuration for a variety of superconductors. In the example given, we predict a significant (4-6×) enhancement in $T_c$ for $MgB_2$, for which $\lambda < \lambda_{max}$. This enhancement may be in fact quite reasonable, as a similar 10× enhancement has been confirmed experimentally in the 2D superconductor FeSe deposited on strontium titanate, due to dielectric environment factors discussed here.

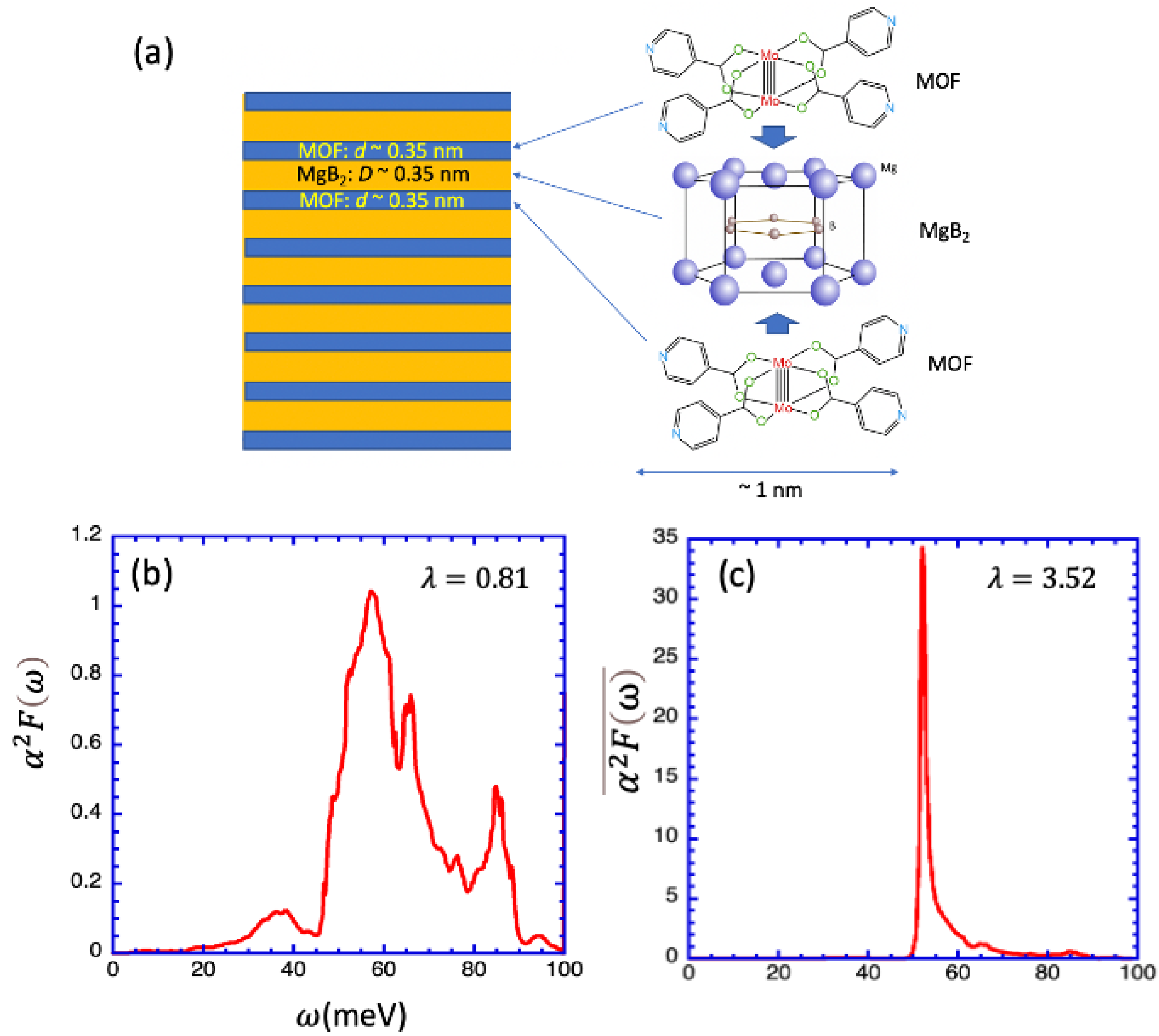

**FIG. 1.** (a) Proposed superlattice arrangement with alternating layers of $MgB_2$ and the MOF material $Mo_2(INA)_4$. The expanded section of a single MOF-$MgB_2$-MOF sandwich shows schematically its structure, with unit cells of MOF and $MgB_2$. (b) Eliashberg function $\alpha^2F(\omega)$ calculated for $MgB_2$ [15,28]. (c) RAS-renormalized $\overline{\alpha^2F(\omega)}$ for $f = 0.2$.